\documentstyle[prl,aps,twocolumn,epsfig]{revtex}

\begin{document}
\title{Mechanical and chemical spinodal instabilities in finite quantum systems}
\author{M. Colonna$^{1}$, Ph. Chomaz$^{2}$ and S. Ayik$^{3}$}
\address{$^{1}$ Laboratori Nazionali del Sud, 
Via S. Sofia 44, I-95123 Catania, Italy \\
$^{2}$ GANIL (DSM-CEA/IN2P3-CNRS),\\
B.P.5027, F-14021 Caen cedex, France\\
$^{3}$ Tennessee Technological University,\\
Cookeville TN38505, USA }
\maketitle

\begin{abstract}
Self consistent quantum approaches are used to study the instabilities of
finite nuclear systems. The frequencies of multipole density fluctuations
are determined as a function of dilution and temperature, for several
isotopes. The
spinodal region of the phase diagrams is determined and it appears
that instabilities are reduced by finite
size effects. The role of surface and volume instabilities is
discussed. It is indicated that the important chemical effects 
associated with mechanical
disruption may lead to isospin fractionation.
\end{abstract}


\hspace{-\parindent}PACS numbers: 21.65.+f, 25.70.Pq, 21.60.Ev\\[1ex]

The dynamics of first order phase transitions is often induced by 
instabilities against fluctuations of the order parameter. For instance,
mechanical instabilities lead to liquid-gas phase transitions and chemical
instabilities induce spinodal decomposition of binary alloys. In 
violent heavy ion collisions, nuclear matter may be quenched in the
coexistence region of the nuclear liquid-gas phase diagram. Then, the
observed abundant fragment formation may take place through a rapid
amplification of spinodal instabilities. New experimental results pleading
in favor of such a spinodal decomposition have recently been reported \cite
{Beanliean,Borderie}. From the theoretical point of view, the spinodal
instabilities in finite systems have been mainly studied within
semi-classical or hydrodynamical framework \cite
{Bertsch,Heiselberg,Norenberg,Colonna,Jacquot1}. However, since the relevant
temperatures are comparable to the shell spacing and the wave numbers of the
unstable modes are of the order of Fermi momentum, quantum effects are
expected to be important, as stressed in \cite{Ayik,Jacquot2}.

In this work, we present a fully quantal investigation of the
spinodal instabilities and related phase diagrams of finite nuclear systems.
We determine frequencies and form factors of the unstable collective modes
of an excited expanded system by linearizing the time dependent
Hartree Fock (TDHF) equation. We carry out applications for Ca and Sn
isotopes. The quantum nature of the system is responsible for many dominant
features such as the fact that the first mode to become unstable is the low
lying octupole vibration. Slightly diluted systems are unstable against low
multipole deformations of the surface, which may be associated with
asymmetric binary or ternary fission processes. We extend our analysis to
charge asymmetric systems and show the importance of chemical effects in 
the spinodal fragmentation process.

In TDHF theory, the evolution of the one-body
density matrix $\tilde{\rho}(t)$ is determined by, 
$i\hbar \partial_{t}\tilde{\rho}(t)=[h[\tilde{\rho}],\tilde{\rho}(t)]$, 
where $h[\tilde{\rho}]={\bf p}^{2}/2m+U[\tilde{\rho}]$ is the mean-field Hamiltonian with $U[\tilde{\rho}]$ as the self-consistent potential. To investigate instabilities encountered
during the evolution of an expanding system, one should study the dynamics 
of the small
deviations $\delta {\tilde{\rho}(t)}$ around the TDHF 
trajectory $\tilde{\rho}(t)$ \cite{Vautherin1}. 
It is more convenient to carry out such an investigation
in the "co-moving frame" of the system, which is described by 
the density matrix, 
$\rho (t)={\rm U}^{\dagger}\left( t\right) ~\tilde{\rho}(t)~{\rm U}\left( t\right)$ where ${\rm U}\left( t\right) =
\exp [\frac{\;}{\;}{\frac{i}{\hbar }}\lambda t{Q}]$
with  ${Q}$ as a one-body operator. For example, in our case $Q$ could be a
suitable constraining operator for preparing the system at low densities and 
$\lambda $ is the associated Lagrange multiplier. Then, the TDHF equation in
the moving frame transforms into, 
\begin{equation}
i\hbar \frac{\partial }{\partial t}\rho (t)=
[{h}(t)\frac{\;}{\;}\lambda {Q},~{\rho (t)}],
\end{equation}
where the mean-field Hamiltonian in the moving frame is given by, 
${h(t)}={\rm U}^{\dagger}\left( t\right) ~h[\tilde{\rho}]~{\rm U}\left( t\right).$ 
The small density fluctuations $\delta \rho (t)$ in the moving frame are determined by the time-dependent RPA equations,
\begin{equation}
i\hbar \frac{\partial }{\partial t}\delta \rho =
[h(t)-\lambda Q,\delta \rho]+[\delta U,\rho ].
\label{EQ:RPA-2}
\end{equation}

Here, we consider the early evolution of instabilities in the vicinity of an
initial state $\rho _{0}$ determined by a constrained Hartree-Fock solution 
$[h(0)-\lambda Q,~\rho _{0}]=0$, where $h(0)=h[{\rho}_0]\equiv h_0 $ is 
the mean-field Hamiltonian at the initial state. Then, small density fluctuations are characterized by the RPA modes $\rho _{\nu }$ and the
associated frequencies $\omega_{\nu }$. 
Incorporating the representation $|i>$, which diagonalizes  $h_{0}-\lambda Q$ 
and $\rho _{0}$ with eigenvalues $\epsilon _{i}$ and occupation 
numbers ${n}_{i}$, 
the equations for the RPA functions $<i|{\rho}_{\nu }|j>={\rho}_{\nu }^{ij}$ become, 
\begin{equation}
\hbar \omega _{\upsilon }\rho _{\nu }^{ij}=(\epsilon _{i}\frac{\;}{\;}%
\epsilon _{j})~\rho _{\nu }^{ij}+\sum_{kl}({n}_{j}\frac{\;}{\;}{n}%
_{i})V_{il,kj}\;\rho _{\nu }^{kl}~,  
\label{omega}
\end{equation}
where $V_{il,kj}=<i|\partial U/\partial \rho _{lk}|j>$ denotes the residual
interaction \cite{Vautherin1,Ring}.
When the frequency of a mode drops to
zero and then becomes imaginary, the system enters an instability region.

To performe an extensive study of instabilities we may
parametrize the possible densities $\rho _{0}$ either by a static
Hartree-Fock (HF) calculations constrained by a set of collective operators 
\cite{Sagawa}, or using a direct parameterization of the density matrix. 
We follow the second approach by introducing a self-similar
scaling of the HF density as suggested by dynamical simulations.

We solve the HF equation for the ground state $\left[ h_{HF},\rho
_{HF}\right] =0$, leading to the single-particle wave functions $|\varphi
_{i}>$ and the associated energies $\varepsilon _{i}$. We introduce the
density matrix at a finite temperature $T$ as $\rho _{HF}\left[ T\right]
=1/\left( 1+\exp \left( \left( h_{HF}\frac{\;}{\;}\varepsilon _{F}\left[
T\right] \right) /T\right) \right) $, where $\varepsilon _{F}\left[ T\right] 
$ is the corresponding Fermi level that is tuned in order to get the correct
particle number. We perform a scaling transformation, $R\left[ \alpha
\right] ,$ which inflates the wave functions in radial direction by a factor 
$\alpha $ according to $<r|R\left[ \alpha \right] |\varphi >=\alpha
^{-1/3}<r/\alpha |\varphi >.$ We define the density matrix for a hot
and diluted system by $\rho _{0}\left[ \alpha ,T\right] =R\left[ \alpha
\right] \;\rho _{HF}\left[ \alpha ^{2}T\right] \;R^{\dagger }\left[ \alpha
\right] $. The associated constrained Hamiltonian is thus $\bar{h}_{0}\left[
\alpha \right] =\alpha ^{2}R\left[ \alpha \right] \;h_{HF}\;R^{\dagger
}\left[ \alpha \right] $ so that the constraint can be identified as $%
-\lambda Q_{\alpha }=\bar{h}_{0}\left[ \alpha \right] -h[\rho _{0}\left[
\alpha ,T\right] ].$ By construction $\left[ \bar{h}_{0}\left[ \alpha
\right] ,~\rho _{0}\left[ \alpha ,T\right] \right] =0$, so that they can be
diagonalized simultaneously. The eigenstates of the constrained Hamiltonian
are given by $|i>=R\left[ \alpha \right] |\varphi _{i}>$, and the
corresponding energies and occupation numbers are $\epsilon
_{i}=\varepsilon _{i}/\alpha ^{2}$ and $n_{i}=1/\left( 1+\exp \left( \left(
\varepsilon _{i}\frac{\;}{\;}\varepsilon _{F}\left[ \alpha ^{2}T\right]
\right) /\alpha ^{2}T\right) \right) $, respectively.
\begin{figure}[tbh]
        \begin{center}
    \epsfig{figure=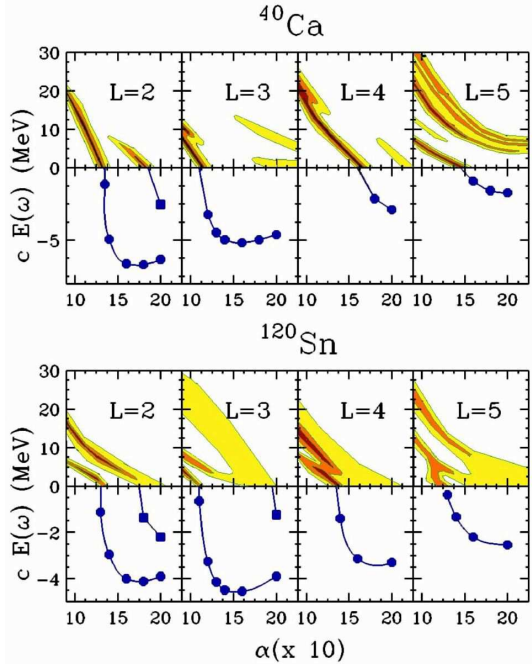,height=9.cm}
     \end{center}
\caption{Contour plots of the isoscalar strength functions associated with
the multipolarity $L=2-5$ as a function of the dilution parameter $\alpha $
and the collective energy of the mode $E_{\nu }=c\hbar \omega _{\nu }$ ($%
c=1$ for stable modes, $-i$ for unstable modes) for $^{40}Ca$ (top) and $%
^{120}$Sn (bottom). 
The strength values are given in percentage of the
EWSR: 2 (yellow), 15 (orange) and 30 (red).}
\end{figure}

We perform the HF\ calculations in the coordinate representation using the
Skyrme force SLy4 \cite{Chabanat}. We note that the isospin symmetry is
already broken at the HF level. The particle and hole states are obtained by
diagonalizing the HF Hamiltonian in a large harmonic oscillator
representation \cite{VanGiai}, which includes 12 major shells for Ca
isotopes and 15  for Sn. We apply the scaling and heating procedures
described above to the density matrix, and calculate the residual
interaction in a self consistent manner. We solve the RPA eq.(\ref{omega})
by a direct diagonalization using a discrete two quasi-particle excitation
representation \cite{Vautherin2}. 

The top part of figure 1 shows calculations performed for $^{40}Ca$. 
Top panels shows contour plots of
the isoscalar strength function associated with the isoscalar operator $%
A_{LM}^{s}=\sum_{i=1}^{A}r_{i}^{L}Y_{LM}$, with multipolarity $L=2-5$, as
a function of the dilution parameter $\alpha $. In the
stable domain, the energy associated with the dominant isoscalar strength
decreases as dilution becomes larger, and at a critical dilution it drops to
zero. At larger dilution, the system becomes unstable, and for each
multipolarity, one or two unstable modes appear. This is illustrated in
the bottom panel, where the ''energy'' of the mode $E_{\nu }=-i\hbar
\omega _{\nu }$ is plotted as a function of the dilution.

It is seen that at low dilutions, around $\alpha =1.2$, only the octupole
mode becomes unstable. In general, density fluctuations with odd
multipolarity become unstable at relatively smaller values of dilution than
those for even multipole fluctuations. This is a genuine quantum effect, due
to the fact that the majority of the particles have to jump only one major
shell to produce an odd natural-parity particle-hole excitation
while twice this energy is required for an even one. In  nuclei at normal
density this makes the first $3^{-}$ a strongly collective low lying state; in
diluted system this state is the first to turn unstable. 
The second feature is that large multipole deformations are hardly
becoming unstable. This is due to surface and quantum effects that prevent
the break-up of such a small system into several fragments. As a
consequence, the fastest growth time, $\tau _{\nu }=\hbar /|E_{\nu }|\approx $
$28$ $fm/c$, occurs for $L=2$ at the dilution $\alpha $=1.8. However, deep
inside the instability region, the octupole mode is almost as unstable as the
quadrupole mode.

Results of similar calculations performed for $^{120}$Sn are shown in the
bottom part of figure 1. Also in this case, the octupole mode becomes
unstable first, at $\alpha \approx 1.1.$
Large multipoles are more unstable in $^{120}$Sn than in $^{40}$Ca, since the
system is larger and can afford larger multipole deformations. This is the
reason why the smallest growth time in $^{120}$Sn occurs for $L=3$.
%

It is useful to study the behavior of the RPA solution in the coordinate
space, $\rho _{p,n}^{+}({\bf r})\equiv \rho _{p,n}^{+}({\bf r},{\bf r})$,
for protons and for neutrons separately. Since the unstable modes have large
isoscalar strength, protons and neutrons mostly move in phase. However,
because of the isospin symmetry breaking induced by the Coulomb force and by
the initial asymmetry of the system considered, neutrons and protons
oscillations have different amplitude and shape. Figure 2(a) shows the
radial dependence of the form factor associated with the unstable octupole
mode, at the dilution $\alpha =1.5$, for protons (dotted line), neutrons
(full line) and the sum (dashed line), in Sn isotopes. Contour
plots of neutron (proton) perturbed densities, ${\rho _{0}}_{p,n}({\bf r}%
)+\rho _{p,n}^{+}({\bf r})$, are also shown in parts b (c) of figure 2. 
\begin{figure}[tbh]
    \begin{center}
    \epsfig{figure=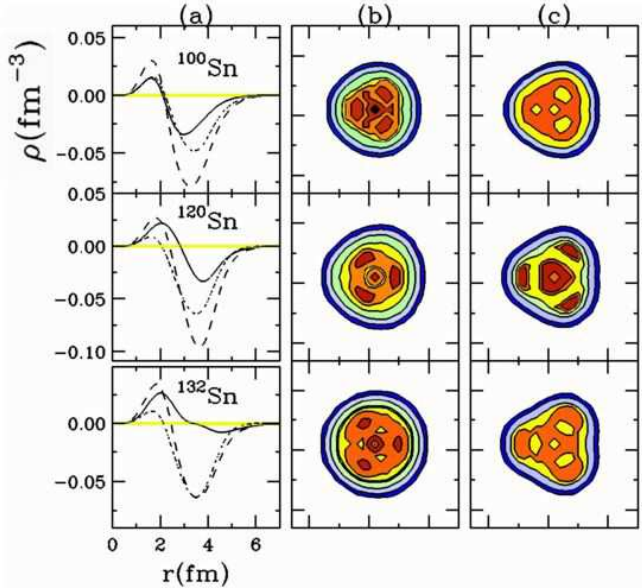,height=8.cm}
     \end{center}
\caption{(a) Radial dependence of the form factor associated with the
unstable mode with L=3, at the dilution $\alpha =1.5$, for protons (dotted
line), neutrons (full line) and the sum (dashed line). The radial distance
is scaled by the dilution parameter $\alpha$.  
(b) Contour plots of
the neutron{\protect\Huge \ }perturbed density. 
From the surface to the center, the contour lines correspond to  
the density values: 0.0037, 0.0075, 0.015, 
and 0.023 $fm^{-3}$. For larger density, colors show small
density variations (3$\%$).
(c) Same for protons}
\end{figure}
We
observe that proton oscillations are mostly located at the surface of the
system, which is a way to minimize the Coulomb repulsion energy. 
At the same time neutrons 
try to follow protons, however this
motion is strongly affected by the isospin initial asymmetry and the
difference between the neutron and proton orbitals at the Fermi energy. 
In fact, in neutron rich systems, much larger proton oscillations are 
observed, that is a way to form more symmetric fragments and hence to reduce
the symmetry energy. 
The
last effect is particularly evident in the neutron rich $^{132}$Sn for which
the neutrons are more difficult to put in motion. 
We observe a quite complex structure of the unstable modes: Volume and
surface instabilities are generally coupled and cannot be easily
disantangled, as well as isoscalar and isovector excitations, since
protons and neutrons do not move in the same way.  
Figure 3 shows the correlations between the amplitude of proton fluctuations 
$\rho _{p}^{+}(r)$ and neutron fluctuations $\rho _{n}^{+}(r)$, obtained
at a radial distance $r$ ranging from $0$ to the system radius $R$, for the octupole mode. 
The different panels correspond to the three
Sn isotopes considered and to three dilutions, $\alpha$, indicated in the 
figure.
 The correlations describe closed paths, since 
$\rho _{p,n}^{+}(r)$ starts from zero at $r=0$ and goes to zero again at
large distances. 
\begin{figure}[tbh]
    \begin{center}
    \epsfig{figure=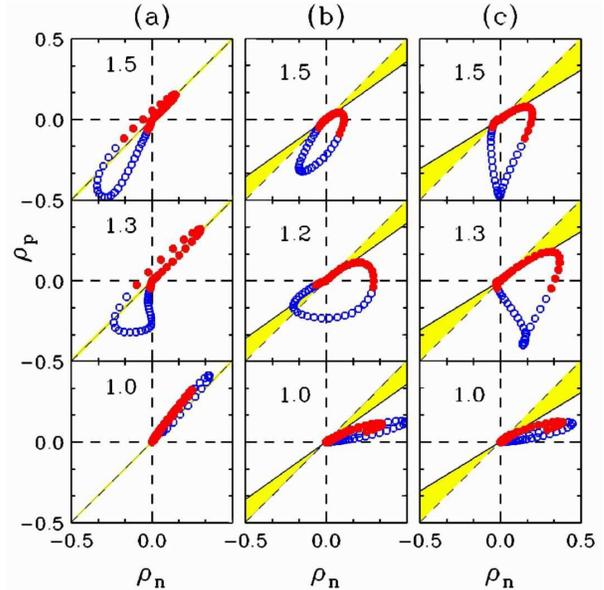,height=8.cm}
     \end{center}
\caption{Correlations between the amplitude of proton and neutron
fluctuations, for several dilutions. The full circles correspond to a radial
distance $0<r<2.4~\alpha~fm$, 
the open circles correspond to $2.4~\alpha~fm<r<R$. The dashed
line indicates $\rho _{p}^{+}=\rho _{n}^{+}$, the full line corresponds to $%
\rho _{p}^{+}=\rho _{n}^{+}~Z/N$. Results are shown for $^{100}$Sn (a), $%
^{120}$Sn (b) and $^{132}$Sn (c). }
\end{figure}
When protons and neutrons fluctuate whithout changing the
local chemical ratio one should have $\rho _{p}^{+}(r)=\rho _{n}^{+}(r)~Z/N$
(full line). An indentical motion of the two fluids should follow the
diagonal $\rho _{p}^{+}(r)=\rho _{n}^{+}(r)$ (dashed line). In $N=Z$ nuclei
(part a) the motion of protons and neutrons should be the same (isoscalar
excitations), however the Coulomb force introduces small differences. As seen
in figure 2, in diluted systems ($\alpha > 1$), we observe
that, in the vicinity of the nuclear surface, density
fluctuations are larger for protons than for neutrons, expecially in the case
of 
neutron-rich unstable isotopes, $^{120}$Sn and $^{132}$Sn 
(see panels (b) and (c) in figure
3), leading to a reduction of the asymmetry of the fragments produced at the
surface. In the interior we observe, for these isotopes, an interesting
evolution
from  $\rho _{p}^{+}(r)<\rho _{n}^{+}(r)~Z/N$ for $\alpha=1$ (stable modes)
to $\rho _{p}^{+}(r)>\rho _{n}^{+}(r)~Z/N$ for the dilute systems. Hence we
observe a change of the local chemical ratio, leading to a reduction of the
symmetry energy, also in the interior of the system.
This effect may be related to the isospin fractionation that occurs in unstable
asymmetric nuclear matter \cite{Virgil}. 
The proton migration towards the large 
density
domains is more frequent than the neutron migration, which may lead to
formation of more symmetric fragments.

We, also, carry out calculations at finite temperature and determine the
dilutions at which different unstable modes begin to appear. This allows us
to specify the border of the instability region in the density-temperature
plane for different unstable modes. Figure 4 shows phase diagrams for
octupole instabilities in Sn and Ca isotopes. 
\begin{figure}[tbh]
    \begin{center}
    \epsfig{figure=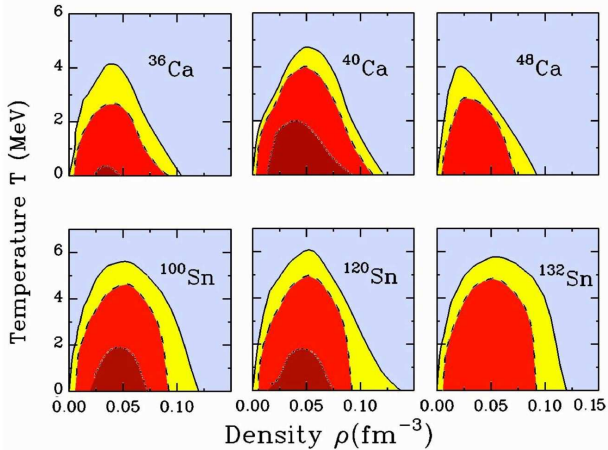,height=6.cm}
     \end{center}
\caption{ Border of the instability region (full fine) associated with $L=3$%
, for $Ca$ and $Sn$ isotopes. The dashed line connects the points having the
instability growth time $\tau $ = 100~fm/c. The dots are associated with $%
\tau $ = 50~fm/c. }
\end{figure}
Here, for simplicity, 
we define the density as $\rho =\rho _{0}/{\alpha }^{3}$. 
The full line 
indicates the border of the 
instability region. 
The dashed line 
connects points that are associated with the instability growth
time $\tau $ = 100~fm/c, and the dots 
corresponds to situations (if any) with a
shorter growth time $\tau $ = 50~fm/c. 
The instability region appears quite reduced as compared to that of nuclear 
matter.
The limiting temperature for instability to occur is around 6 MeV for Sn and
4.5 MeV for Ca while it is about $16~MeV$ in symmetric nuclear matter. 
Heavier systems have larger instability region than the lighter ones.
Moreover, more asymmetric systems are less unstable. In fact, in spite of
the larger mass, $^{132}$Sn is less unstable than $^{120}$Sn. Larger
instability growth times are obtained for $^{132}$Sn. As seen from phase
diagrams for $^{36}$Ca and $^{40}$Ca, the region of instability is reduced
also in proton-rich systems. This behavior is in agreement with nuclear
matter calculations, which indicates that the instability region shrinks in
asymmetric nuclear matter \cite{Virgil}.

In conclusion, we have presented a study of the early development of
spinodal instabilities in finite nuclear systems by employing a quantal RPA
approach. Results are relevant for multifragmentation studies, in fact
dominant unstable modes determine the onset of the subsequent
fragmentation of the system, that has to be followed using a full
dynamical calculation.  
We have investigated isospin effects on
instabilities by carrying out calculations for Ca and Sn isotopes. 
We find that instabilities are mostly of isoscalar nature, but contain also
an important isovector component in asymmetric matter. Hence the
liquid-gas separation of asymmetric systems is always linked to a  
chemical separation inducing a fractional distillation of the system.
The degree of instability in neutron-rich and also
proton-rich nuclei appears reduced as compared to that of symmetric nuclei of
comparable size. The instabilities are also reduced in small nuclei.
Finally we have stressed important quantum effects, 
such as the octupolar
nature of the most important instability.\\[1ex]  

Authors acknowledge the Yukawa Institute for Theoretical Physics,
S.A. and M.C. acknowledge GANIL, and Ph.C. acknowledges LNS, where parts of
this work have been carried out, for support and warm hospitality. 
Stimulating discussions with M.Di Toro and D.Lacroix are also
acknowledged.
This work
is supported in part by the US DOE grant DE-FG05-89ER40530. 

\end{document}